\documentclass{ws-mpla}
\usepackage[super]{cite}
\usepackage{graphicx}
\begin{document}

\markboth{Marc Sher}{Flavor-changing neutral currents in the Higgs sector}

\catchline{}{}{}{}{}

\title{Flavor Changing Neutral Currents\\
in the Higgs Sector}

\author{Marc Sher}

\address{Physics Department, William \& Mary\\
Williamsburg, VA  23187, USA
\\
mtsher@wm.edu}

\maketitle

\pub{Received (Day Month Year)}{Revised (Day Month Year)}

\begin{abstract}
This review considers models with extended Higgs sectors in which there are tree-level flavor-changing neutral currents (FCNC) mediated by scalars.   After briefly reviewing models without tree-level FCNC, several models with such currents are discussed.     A popular mass-matrix ansatz, in which the flavor-changing couplings are the geometric mean of the individual flavor couplings, is presented.   While it provided a target for experimentalists for three decades, it in now being severely challenged by experiments.   Couplings expected to be of O(1) must be substantially smaller and the ansatz is now not favored.    The minimal flavor violation hypothesis is introduced.  Then specific models are presented, including the Branco-Grimus-Lavoura models.   These models are not yet excluded experimentally, but they are highly predictive and will be tested once heavy Higgs bosons are discovered.   We then turn to flavorful models and flavor-changing decays of heavy Higgs bosons, and it is shown that in many of these models, the heavy Higgs could predominantly decay in a flavor-changing manner (such as $ct$ or $\mu\tau$) and experimentalists are encouraged to include these possibilities in their searches.
\end{abstract}

\ccode{12.15.Mm,12.60.-i,14.80.Cp}

\section{Introduction}	

Flavor-Changing Neutral Currents (FCNC) played a critical role in the development of the Standard Model and continue to be intensely studied.
The Standard Model originally had three quarks, $u, d$ and s, and this led to FCNC processes that occurred far too rapidly.   To remedy this, the GIM mechanism
\cite{Glashow:1970gm} introduced the c quark, thus removing the dangerous tree level FCNC.    At one loop, of course, FCNC will occur.  This fact was used by Gaillard and Lee\cite{Gaillard:1974hs} to successfully predict the c quark mass.   This pattern was repeated with the third generation (although the top quark mass prediction was less accurate due to uncertainties in the CKM mixing angles).    It is now well established that the neutral gauge boson couplings in the Standard Model are flavor diagonal.

The same is true for the Higgs sector of the Standard Model (SM).    The most general Yukawa Lagrangian of the SM immediately yields, upon spontaneous symmetry breaking, a mass matrix which is proportional to the Yukawa coupling matrix
$${\cal L}_Y = y_{ij}\bar{\psi}_i\psi_j \Phi \Longrightarrow  M_{ij} = \frac{y_{ij}}{\sqrt{2}} \langle \phi \rangle.$$
Here, the subscripts are generation indices.    One can see that diagonalizing the mass matrix automatically diagonalizes the Yukawa coupling matrix and thus the Higgs couplings to fermions are also flavor-diagonal.

Some of the most studied extensions of the SM involve extensions of the scalar sector.     Models of electroweak baryogenesis require such extensions (Refs. \refcite{Bochkarev:1990fx,Turok:1990zg,Land:1992sm,Cline:1995dg,Cline:1996mga} are some of the original papers, a recent work with a comprehensive list of references is Ref. \refcite{Anisha:2022hgv}) and supersymmetric models require additional doublets.    Some extensions, such as the inert doublet model, lead to an attractive dark matter candidate.   A recent review of various models with an extremely extensive list of references can be found in Ref. \refcite{Ivanov:2017dad}.    In this review, I will focus on models with two Higgs doublets (2HDMs) and look at the issue of FCNC in these models.    A detailed review article on 2HDMs can be found in Ref. \refcite{Branco:2011iw}.

In a 2HDM, the most general Yukawa couplings are 
$${\cal L}_Y = y^1_{ij}\bar{\psi}_i\psi_j \Phi_1 +  y^2_{ij}\bar{\psi}_i\psi_j \Phi _2$$
which leads to the mass matrix
$$ M_{ij} = y^1_{ij} \frac{v_1}{\sqrt{2}} + y^2_{ij} \frac{v_2}{\sqrt{2}}$$
where $v_1$ and $v_2$ are the vacuum expectation values of the two Higgs doublets.    Since $y^1$ and $y^2$ are, in general, not simultaneously diagonalizable, this will lead to tree level FCNC.    These FCNC are very problematic.   The $\bar{d}s\phi$ coupling, for instance, will lead to large $K-\bar{K}$ mixing unless the coupling is very small, the $\phi$ is very heavy or there is a cancellation between the contributions of different scalars.  If the coupling is $f\bar{d}s\phi$, then the lower bound to the $\phi$ mass would be approximately $7000 f $ TeV.  So $f$ needs to be very small to be acceptable.   Other bounds, as we will see later, involve $B-\bar{B}$, $B_s-\bar{B}_s$. $D-\bar{D}$ mixing as well as rare B decays, tau and muon decays, etc.

There are two approaches to solving this problem.   In the first, a discrete symmetry is imposed which eliminates the problematic terms.   In the second, some principle is used to ensure that the couplings are sufficiently small.    In the next section, the models with a discrete symmetry will be reviewed.   In the following section, models without such a symmetry will be discussed.

\section{Models without tree-level FCNC}

In the 2HDM, there are two Higgs doublets with hypercharge\footnote{An alternative convention is for the second doublet to have hypercharge $-1$, in which $\Phi_2 \rightarrow
i\epsilon\Phi_2^*$}\ $Y=+1$
$$\Phi_1 = \begin{pmatrix} \phi_1^+\\ \phi_1^0\end{pmatrix} \qquad \Phi_2 = \begin{pmatrix} \phi_2^+\\ \phi_2^0\end{pmatrix}$$
Of the eight fields, three are eaten to become the longitudinal components of the $W^\pm$ and $Z$ and the remaining five consist of a charged scalar, $H^\pm$, a pseudoscalar $A$ and two neutral scalars, $H$ and $h$.   In general, the masses depend on somewhat arbitrary scalar self-couplings.

As noted above, the general 2HDM has tree-level FCNC.   The most attractive way to eliminate these currents is to impose a discrete symmetry.    The Glashow-Weinberg theorem\cite{Paschos:1976ay,Glashow:1976nt}  states that in models with Higgs doublets and singlets, tree-level FCNC will be eliminated if all quarks of a given charge couple to only one Higgs doublet.    It is often stated that the theorem applies to all {\it fermions} of given charge and thus it is also applied to the lepton sector as well, however (as Glashow and Weinberg originally stated) it need not necessarily apply to the lepton sector; this will be discussed shortly.

Using a simple $Z_2$ symmetry $\Phi_1 \leftrightarrow -\Phi_1$, one gets the Type I 2HDM in which all fermions couple to one doublet, $\Phi_2$.     The Type II 2HDM is obtained by imposing $\Phi_2 \leftrightarrow -\Phi_2, u_R \leftrightarrow -u_R$, in which case the $Q=2/3$ quarks couple to $\Phi_2$ and the $Q=-1/3$ quarks and the leptons couple to $\Phi_1$.   The Type II model is the most studied, since axion and supersymmetric models are Type II.   Two alternatives (the lepton-specific and flipped models) can be obtained by having the leptons and $Q=-1/3$ quarks couple to different multiplets.  These are summarized in Table 1.

\begin{table}[h]
\tbl{The most familiar 2HDMs.}
{\begin{tabular}{@{}cccc@{}} \toprule
Model & $u_R^i$ & $d_R^i$ &
$e_R^i$ \\
\colrule
Type I& $\Phi_2$ &$ \Phi_2$ &$ \Phi_2$ \\
Type II&$ \Phi_2$ &$ \Phi_1$ & $\Phi_1$ \\
Lepton-specific & $\Phi_2$ &$ \Phi_2$ & $\Phi_1$ \\
Flipped &$ \Phi_2$ & $\Phi_1$&$ \Phi_2$\\ \botrule
\end{tabular}\label{ta1} }
\end{table}
The two doublets can be written as 
\begin{equation}
\Phi_j = \begin{pmatrix} \phi_j^+ \\ (v_j + \rho_j + i\eta_j)/\sqrt{2}\end{pmatrix}.
\end{equation}    
We define $\tan\beta \equiv v_2/v_1$ and then the pseudoscalar is $A = \eta_1\sin\beta - \eta_2\cos\beta$.   The neutral scalars depend on another parameter $\alpha$ which gives $h = \rho_1\sin\alpha + \rho_2\cos\alpha$ and $H = \rho_1\cos\alpha -\rho_2\sin\alpha$.   The Yukawa couplings relative to the Standard Model couplings then depend on $\alpha$ and $\beta$ and can be written as
\begin{align}
{\cal L}^{2HDM}_{Yukawa} &= 
-\sum_{f=u,d,\ell}\frac{m_f}{v} \left( \xi_h^f \bar{f}f h + \xi_H^f \bar{f}fH -i\xi_A^f\bar{f}\gamma_5f\right) \\
&+\left(\frac{\sqrt{2}V_{ud}}{v} \bar{u}(m_u \xi_A^uP_L + m_d\xi_A^dP_R)dH^+ + \frac{\sqrt{2}m_\ell \xi_A^\ell}{v} \bar{\nu}_L \ell_R H^+ + {\rm H.c.}\right)
\end{align}
where the $\xi$ are given in Table 2.   The coupling to the $W$ and $Z$ are the same as in the Standard Model times $\sin(\alpha-\beta)$ for $h$ and $\cos(\alpha-\beta)$ for $H$.   Note that in the ``alignment limit" of $\cos(\alpha-\beta) = 0$, then the couplings of the h to all fields is identical to the Standard Model.

\begin{table}[h]
\tbl{The Yukawa couplings relative to the Standard Model.}
{\begin{tabular}{@{}c|c|c|c|c@{}} \hline\hline
 & Type I & Type II &
Lepton-specific & Flipped \\
\hline
$\xi^u_h$ & $\cos\alpha/\sin\beta$ &$ \cos\alpha/\sin\beta$ & $\cos\alpha/\sin\beta$ & $\cos\alpha/\sin\beta$ \\
\hline
$\xi^d_h$& $ \cos\alpha/\sin\beta$ & $-\sin\alpha/\cos\beta$ & $ \cos\alpha/\sin\beta $ & $-\sin\alpha/\cos\beta$ \\
\hline
$\xi^\ell_h$ & $ \cos\alpha/\sin\beta$ & $-\sin\alpha/\cos\beta$ & $ -\sin\alpha/\cos\beta $ & $\cos\alpha/\sin\beta$ \\
\hline
$\xi^u_H$ & $ \sin\alpha/\sin\beta$ & $\sin\alpha/\sin\beta$ & $ \sin\alpha/\sin\beta $ & $-\sin\alpha/\sin\beta$ \\
\hline
$\xi^d_H$ & $ \sin\alpha/\sin\beta$ & $\cos\alpha/\cos\beta$ & $ \sin\alpha/\sin\beta $ & $\cos\alpha/\cos\beta$ \\
\hline
$\xi^\ell_H$ & $ \sin\alpha/\sin\beta$ & $\cos\alpha/\cos\beta$ & $ \cos\alpha/\cos\beta $ & $\sin\alpha/\sin\beta$ \\
\hline
$\xi^u_A$ & $\cot\beta$ & $\cot\beta$ & $ \cot\beta$ & $\cot\beta$ \\
\hline
$\xi^d_A$ & $-\cot\beta$ & $\tan\beta$ & $ -\cot\beta$ & $\tan\beta$ \\
\hline
$\xi^\ell_A$ & $-\cot\beta$& $\tan\beta$ & $ \tan\beta$ & $-\cot\beta$ \\
\hline\hline
\end{tabular}\label{ta2} }
\end{table}

There are dozens of papers which plot the allowed region in the $\tan\beta - \cos(\beta-\alpha)$ plane (see  Figure 3 of Ref. \refcite{Robens:2022oue} for the most recent at the time of this writing).   Typically, the Type II and Flipped models require $\cos(\beta-\alpha)$ to be quite small (less than $0.1$), whereas the Type I and Lepton-specific models allow substantially larger $\cos(\beta-\alpha)$.
For completeness, the inert doublet model should be mentioned. Here the Z2 symmetry, which usually is softly broken in the above four models, is unbroken. The additional scalar does not get a vacuum expectation value and does not couple to fermions. Thus the lightest additional scalar is absolutely stable and is an excellent candidate for dark matter. A  review of the model with an extensive list of references can be found in Ref. \refcite{Ilnicka:2018def}. Since we are more concerned with models with tree-level FCNC in this review, we will not focus on these models further.

As noted above, there is a loophole to the commonly stated version of the Glashow-Weinberg theorem.    It is impossible for the up and down quark mass matrices to be diagonal since that would eliminate CKM mixing.   However, the charged lepton and Dirac neutrino mass matrices can be diagonal, since the PMNS mixing could come (via a conventional seesaw) from the superheavy Majorana neutrino sector.     Abe, Sato and Yagyu (ASY) \cite{Abe:2017jqo} constructed a model in which the muon coupled to $\Phi_1$ and the other fermions coupled to $\Phi_2$.    This was implemented in ASY by imposing a $Z_4$ symmetry, but it was shown by Ivanov and Nishi\cite{Ivanov:2013bka} that the actual symmetry is $Z_2 \times U(1)$ - the $Z_2$ is just $\Phi_1 \leftrightarrow -\Phi_1,\ \mu_R \leftrightarrow -\mu_R$ and the $U(1)$ is muon number.    This causes the muon and muon neutrino parts of the mass matrices to decouple from the $e$ and $\tau$ parts, but as noted above, this does not eliminate muon neutrino mass mixing due to superheavy Majorana neutrino mixing.   The purpose of the ASY study was to explain the anomalous magnetic moment of the muon.   Ferreira and MS \cite{Ferreira:2020ukv} later argued that the explanation  of the magnetic moment required extensive fine-tuning.   They performed a detailed phenomenological analysis of the model, showing that the Higgs dimuon coupling can be substantially altered.    They found that $\xi_\mu = \frac{1-\xi_Z\xi_\tau}{\xi_Z - \xi_\tau}$ where  $\xi_f$ is the ratio of the Higgs coupling of $f$ to the SM value.   They also studied the charged Higgs phenomenology, showing the $H^+ \rightarrow \mu\nu_\mu$ can be the dominant decay, and also looked at the phenomenology of the heavy neutral scalars in the model.

We now turn to models that contain tree-level FCNC.

\section{Models with tree-level FCNC}

If FCNC exist at tree level, then some ansatz or principle must be invoked to make the flavor-changing couplings sufficiently small.   There are two general schemes that have attracted wide attention.  The first is the so-called Type III 2HDM with a mass-matrix ansatz  (the so-called\footnote{The name was not proposed by the author, but appeared in early work of Hou and others.} ``Cheng-Sher ansatz") and the second is minimal flavor violation.   Although I was one of the originators of the former scheme, I believe that the second scheme is more compelling and more consistent with recent experimental results.    More recently, flavorful models have been developed.    Each will now be discussed, followed by some comments about FCNC involving heavy Higgs bosons.

\subsection{The rise and fall of the Cheng-Sher ansatz}

In discussing FCNC, it is more convenient to rotate to the Higgs basis in which one field gets a vev and the other does not.   In that case
\begin{align}
{\cal L}_{\rm Yukawa} &= \eta_{ij}^u \bar{Q}_{iL}\tilde{H}_1 U_{jR} + \eta_{ij}^d \bar{Q}_{iL}H_1 D_{jR} + \eta_{ij}^\ell \bar{L}_{iL}H_1 E_{jR}\\
&+  \xi_{ij}^u \bar{Q}_{iL}\tilde{H}_2 U_{jR} + \xi_{ij}^d \bar{Q}_{iL}H_2 D_{jR} + \xi_{ij}^\ell \bar{L}_{iL}H_2 E_{jR}\end{align}
where
\begin{equation}
\langle H_1 \rangle_0 = \begin{pmatrix} 0\\v/\sqrt{2} \end{pmatrix}\quad , \quad \langle H_2 \rangle_0 = \begin{pmatrix} 0\\0\end{pmatrix}
\end{equation}
Diagonalizing the mass matrix does diagonalize the $\eta_{ij}$ couplings, but NOT the $\xi_{ij}$ couplings, leading to tree level FCNC.

Although it was known in the late 70's that a $Z_2$ symmetry avoided these FCNC, the introduction of such a symmetry seemed ad hoc, and thus it was questioned as to how necessary it was.     In 1980, some experimenters looking at $K_L \rightarrow \mu e$ said that ``if the flavor-changing coupling is O(1), we find a lower bound on the Higgs mass of 60 TeV - this is higher than the energy of the SSC!".    Of course, this ignored mixing, the difference between the two Higgs bosons, etc., but this does demonstrate the extreme sensitivity to the value of the flavor-changing coupling.    A more realistic assumption was made by McWilliams and Li \cite{McWilliams:1980kj} and by Shanker \cite{Shanker:1981mj} who assumed that the flavor-changing coupling was the heaviest fermion of that particular charge times a mixing angle.   Since the angle was unknown, they assumed it was O(1).    That still gave a bound on a Higgs mass of a few TeV from $K_L \rightarrow \mu e$ and an even higher bound of 100 TeV from $\Delta m_K$ (although there are greater uncertainties).    Partly for these reasons (and the rise of supersymmetry which gave the Type II structure automatically), FCNC at tree level was generally ignored for most of the decade.

In the mid-80's, there was great interest in Fritzsch type matrices  $\begin{pmatrix} 0 & A \\ A& B \end{pmatrix}$ which, if $A << B$, has eigenvalues of $A^2/B$ and $B$.  The off-diagonal term is the geometric mean of the eigenvalues.   If this is the down quark mass matrix, this leads to the numerically correct result that $\sin\theta_c = \sqrt{m_d/m_s}$.   Cheng and Sher \cite{Cheng:1987rs} (CS) proposed that the flavor-changing couplings should be of the order of the geometric mean of the Yukawa couplings of the two fermions, i.e.
\begin{equation}
\xi_{ij} = \lambda_{ij} \sqrt{m_im_j}\frac{\sqrt{2}}{v}
\end{equation}
where the $\lambda_{ij}$ are of order one.   This substantially reduces the Yukawa couplings involving the first two generations, alleviating the bounds of the previous paragraph.

Their argument was based on the following:   If the fermion mass matrix is Fritzsch-like
\begin{equation} M = \begin{pmatrix}0&A&0\\A&0&B\\0&B&C\end{pmatrix}\end{equation} then the eigenvalues are approximately given by $A \simeq \sqrt{m_1m_2}, B \simeq  \sqrt{m_2m_3}, C \simeq m_3$.   CS then just assumed that the Yukawa coupling matrices were of the same structure.    Although the Fritzsch-like matrices no longer work well, CS pointed out that their argument only requires that the hierarchy of eigenvalues does not arise through delicate cancellations.   It was later pointed out \cite{Antaramian:1992ya} that if the hierarchical structure is due to approximate flavor symmetries, then the CS ansatz will be satisfied.

The CS ansatz received very little attention for a few years.  Then the top quark turned out to be heavier than most expected and the B-factories (BELLE/BABAR) began.  The ansatz gave experimenters a target - they could give bounds in terms of $\lambda_{ij}$ instead of a generic coupling whose value was arbitrary.  It also meant that B decays and mixings would have a huge increase in precision and thus $\lambda_{ij} = 1$ was in reach.    In 1996, a very comprehensive analysis of the model by Atwood, Reina and Soni \cite{Atwood:1996vj} looked at a large variety of processes including effects on $\Delta m_B$, $t \to c h$, rare $\mu, \tau$ and $B$ decays, $b \to s \gamma$, rate top decays and flavor-changing Z decays.  The review article of Branco, et al \cite{Branco:2011iw} noted earlier gives a detailed list of the various constraints (with several dozen references on this topic) as of 2012.

Since 2012, the bounds on the $\lambda_{ij}$ have become much more precise.    Lattice gauge theory calculations have given us much better information about the various hadronic uncertainties, Higgs decays have now been measured and BELLE/LHCb have provided numerous results.    A very recent analysis of all of these bounds in the context of the CS ansatz is in the work of Babu and Jana \cite{Babu:2018uik}.     They studied the various meson-antimeson constraints and produced Table 3.

\begin{table}[h]
\tbl{Bounds on the $\lambda_{ij}$ from meson mixing, from Babu and Jana\cite{Babu:2018uik}. }
{\begin{tabular}{@{}|c|c|@{}} \hline
  Process & Upper bound on $\lambda_{ij}$ \\
\hline
$K^0 - \overline{K}^0$ mixing & 0.26 \\
\hline
$B_s^0 - \overline{B_s}^0$ mixing & 0.436\\
\hline
$B^0 - \overline{B}^0$ mixing & 0.379\\
\hline
$D^0 - \overline{D}^0$ mixing & 0.222\\
\hline
\end{tabular}\label{ta3} }
\end{table}

These bounds assumed a pseudoscalar mass of $500$ GeV (the bound scales approximately linearly).    The bounds from scalar exchange is roughly a factor of 3 weaker.    Unless the pseudoscalar mass is well above a TeV, these bounds will all be less than 1.0.

\begin{figure}[ph]
\centerline{\includegraphics[width=3.0in]{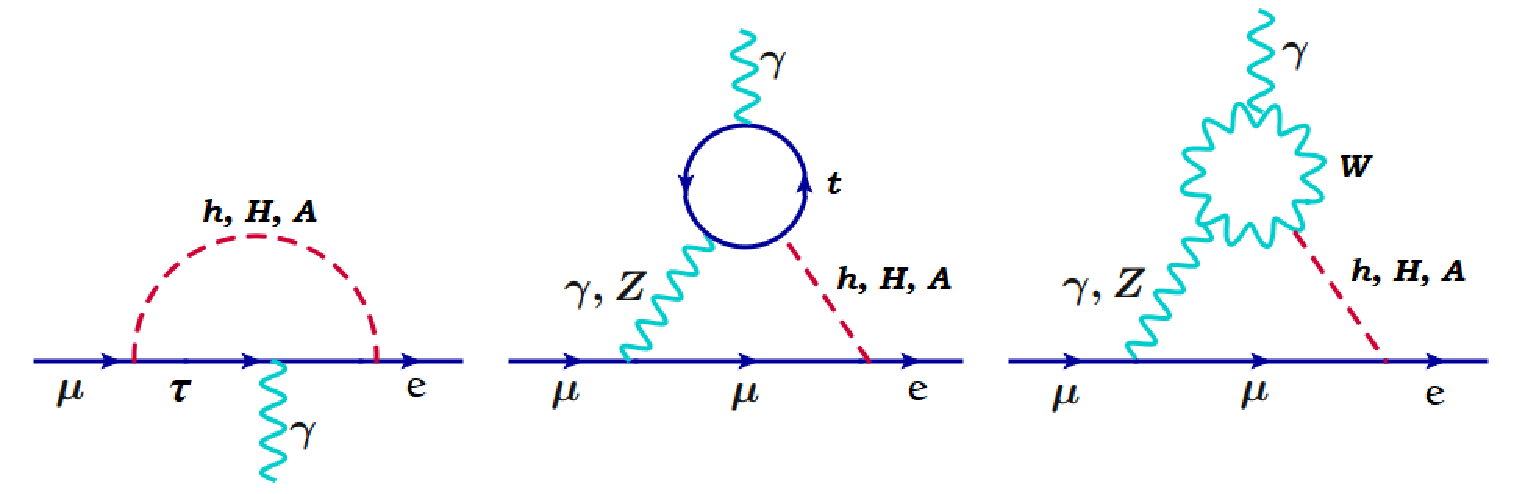}}
\vspace*{6pt}
\caption{One and two loop diagrams contributing to $\mu\to e\gamma$. This figure comes from Ref. 23.\protect\label{fig1}}
\end{figure}

\begin{figure}[h!]
\centerline{\includegraphics[width=3.0in]{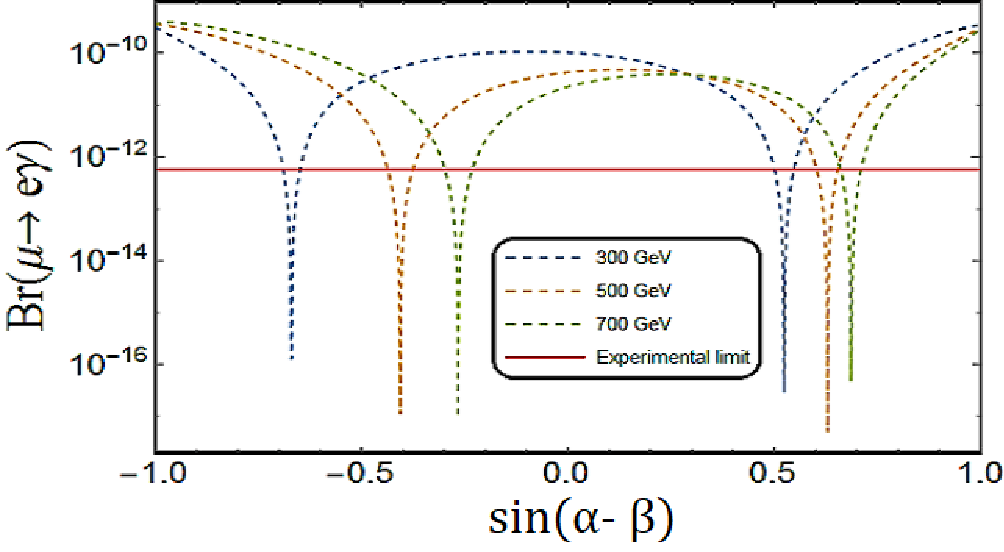}}
\vspace*{6pt}
\caption{Value of the branching ratio for $\mu\to e\gamma$ assuming $\lambda_{e\mu} = 1$ for various scalar masses.   The two-loop diagrams of Figure 1 dominate and the dips occurs due to destructive interference between the top and $W$ loops.   The horizontal line is the current experimental bound.   The figure comes from Ref. 23.   The notation in that paper is different - the term $\sin(\alpha-\beta)$ on the x-axis should be $\cos(\alpha-\beta)$ to match the conventional notation of section 2.   Current limits on the x-axis are typically between $-0.4$ and $0.4$.\protect\label{fig2}}
\end{figure}

One can also study $\mu \rightarrow e\gamma$ in the model.  In this case, the two-loop Barr-Zee diagrams \cite{Barr:1990vd} of Figure 1 can dominate over the one-loop due to a chiral enhancement as first shown by Chang, Hou and Keung \cite{Chang:1993kw}.    More recently, Babu and Jana \cite{Babu:2018uik} studied the model.    They calculate the branching ratio for $\mu \to e\gamma$ as a function of $\sin(\alpha-\beta)$.    Their result is in Figure 2, one sees that with the exception of a couple of very narrow regions in which the top and $W$ loops destructively interfere, $\lambda_{e\mu}$ must be considerably less than one.   In fact, for most of parameter space, it must be less than $0.12$.

What about $\lambda_{\mu\tau}$?    One can look at the decay of the Higgs into $\mu\tau$.  The branching ratio\cite{Sher:2016rhh,Hou:2019grj} is $0.0076\lambda^2_{\mu\tau}\cos^2(\alpha-\beta)$ Using the current CMS experimental bound on the branching ratio of $0.0025$, one finds that $\lambda_{\mu\tau} < 0.6/\cos(\alpha-\beta)$, which is very weak and not very restrictive.   It should be noted that a heavy Higgs boson decaying similarly will have the factor $\cos^2(\alpha-\beta)$ replaced with $\sin^2(\alpha-\beta)$, and this might be much larger.    An analysis of the $\tau\to\mu\gamma$ decay in the model is in Ref. \refcite{Vicente:2019ykr}; the result is that current bounds are beginning to constrain $\lambda_{\mu\tau}$.

Is there any way to avoid these bounds (without fine-tuning)?     One could choose the $v$ in the definition of the $\lambda_{ij}$ to be $v_1$ or $v_2$, i.e. rescaling by a factor of $\cos\beta$, but that violates the spirit of the CS ansatz.    It appears that the CS ansatz is in trouble - five of the nine off-diagonal coefficients, which should be $O(1)$, are substantially smaller.    It is certainly possible that there is some wiggle-room.   The pseudoscalar mass could be well above 1 TeV, the region of parameter space in which destructive interference in $\mu\to e\gamma$ could occur, or there could be a little fine-tuning.    However the ansatz does not appear to be as viable as it once was.   It may still be useful in parametrizing and comparing FCNC studies.

\subsection{Minimal Flavor Violation}
 
 A much more robust solution to the problem of tree-level FCNC is minimal flavor violation (MFV).     This essentially posits that all flavor violating and CP violating interactions are linked to the CKM or PMNS structure of the Yukawa couplings.     The concept first arose in the work of Chivukula and Georgi \cite{Chivukula:1987py} in the context of composite technicolor models and by Hall and Randall \cite{Hall:1990ac} in the context of supersymmetry.    It continued to be assumed for many supersymmetric models over the next decade.    Buras et al. \cite{Buras:2000dm} extended the idea to a 2HDM, but that still was restricted to particular models.   A more formalized description in a general effective field theory approach was given by D'Ambrosio et al \cite{DAmbrosio:2002vsn}.    Blanke et al. \cite{Blanke:2006ig} used the effective field theory description to find model-independent tests of MFV.    A generalization to multi-doublet models was discussed by Botella, et al. \cite{Botella:2009pq}
 
 The basic idea of MFV (see Ref. \refcite{Isidori:2010kg} for a nice discussion) is as follows.   In the absence of Yukawa couplings (concentrating on the quark sector), the Lagrangian has an $SU(3)_{Q_L} \times SU(3)_{U_R} \times SU(3)_{D_R}$ symmetry.    The Yukawa interactions break this symmetry\footnote{A very early paper pointing this out and looking to flavor symmetries to explain the CKM matrix is in Ref. \refcite{Gerard:1982mm}}.    To formally retain flavor symmetry, one can introduce dimensionless auxilliary fields $Y^u$ and $Y^d$ which transform under the above symmetry as $(3,\overline{3},1)$ and $(3,1, \overline{3})$, respectively..    An effective theory satisfies the criterion of MFV in the quark sector if all higher-dimensional operators constructed from the Standard Model and Y fields are invariant under the $SU(3)_{Q_L} \times SU(3)_{U_R} \times SU(3)_{D_R}$ symmetry.   In some versions, invariance under CP is also imposed.    Since most of the Yukawa matrices have very small eigenvalues and since the off-diagonal elements of the CKM matrix are also small, one does not have to go to very high order in writing down effective operators.    It is not as easy to define MFV in the lepton sector since the origin of the PMNS matrix might be related to superheavy Majorana neutrino masses.    The MFV ansatz is not a ``theory of flavor" since there is no explanation of the hierarchical structure, but it is certainly testable in future experiments.    Table III of Ref. \refcite{Isidori:2010kg} lists several potential experimental results that could refute MFV.
 
 A full analysis would be beyond the scope of this review.  In fact, the citation count of the D'Ambrosio et al \cite{DAmbrosio:2002vsn} paper shows the robustness of the MFV hypothesis and the interest in the community.    One should mention a version of MFV without tree level FCNC.    As discussed by Buras et al. \cite{Buras:2010mh}, one can satisfy the MFV hypothesis by assuming that the Yukawa coupling matrices $y_1$ and $y_2$ are proportional \cite{Pich:2009sp}.     The Buras et al. paper does compare this model with the natural flavor conservation models (type I and type II).    However, this review is focusing on tree-level FCNC and thus this model will not be discussed further.     Rather, I will focus on a particular implementation (or UV completion) of the MFV hypothesis known as the BGL model.
 
 In 1996, following an earlier suggestion by Lavoura\cite{Lavoura:1994ty}, Branco, Grimus and Lavoura (BGL) constructed \cite{Branco:1996bq} models in which the FCNC couplings depend only on the elements of the CKM matrix (several years before the phrase ``minimal flavor violation" was used).   They accomplished this by the use of discrete symmetries.   To show how this works, one can just focus on neutral currents involving $Q=-1/3$ quarks.    Taking the vevs to be real for simplicity (BGL were more general), the down quark mass matrix is (as noted earlier) 
 \begin{equation}
 M_d = \frac{1}{\sqrt{2}} (v_1 Y_1 + v_2 Y_2)
 \end{equation}
 This matrix is bi-diagonalized as $U^\dagger_{dL} M_d U_{dR} = D_d \equiv {\rm diag} (m_d,m_s,m_d)$.    In terms of quark mass eigenstates, the flavor-changing neutral currents involving the two neutral scalars and the pseudoscalar are controlled by the matrix $N_d$ which is given by
 \begin{equation}
 N_d = \frac{1}{\sqrt{2}} U^\dagger_{dL} (v_2 Y_1 - v_1 Y_2) U_{dR}
 \end{equation}
 This matrix will in general not be diagonal.    It can be rewritten as
 \begin{equation}
 N_d = \frac{v_2}{v_1} D_d - \frac{v_2}{\sqrt{2}}\left( \frac{v_2}{v_1}+\frac{v_1}{v_2}\right) U^\dagger_{dL} Y_2 U_{dR}
 \end{equation}
 A similar expression will be given for the up-quarks.    Clearly, the second term will not conserve flavor, leading to FCNC.    The CKM matrix is given by $V = U^\dagger_{uL}U_{dL}$ which is clearly different from the last term in Eq. (11).    One needs to get rid of the dependence of $U_{dR}$ and to relate $U^\dagger_{dL}$ to $V$.    
 
BGL showed that a discrete symmetry of the form:
\begin{equation}
Q_{3L} \to e^{i\psi} Q_{3L}, \qquad   u_{3R} \to e^{2i\psi} u_{3R}, \qquad \Phi_2 \to e^{i\psi} \Phi_2
\end{equation}
automatically gives Yukawa matrix textures of the form
\begin{equation}
Y_1^d = \begin{pmatrix} x&x&x\\ x&x&x \\ 0&0&0\end{pmatrix}, \qquad Y_2^d = \begin{pmatrix} 0&0&0 \\ 0&0&0 \\ x&x&x \end{pmatrix}
\end{equation}
\begin{equation}
Y_1^u = \begin{pmatrix} x&x&0\\ x&x&0 \\ 0&0&0\end{pmatrix}, \qquad Y_2^u = \begin{pmatrix} 0&0&0 \\ 0&0&0 \\ 0&0&x \end{pmatrix}
\end{equation}

Plugging these in automatically gives the relation (no sum on j)
\begin{equation}
(N_d)_{ij} = \frac{v_2}{v_1} (D_d)_{ij} - \left( \frac{v_2}{v_1} + \frac{v_1}{v_2}\right) V^\dagger_{i3}V_{3j}(D_d)_{jj}
\end{equation}
This is precisely what is needed - the FCNC only depend on the CKM elements.    Note also that is only depends on the ratios of vevs and is thus very predictive.   In the quark sector, there are six BGL models (one can replace the $``3"$ with a $``2"$ or a $``1"$ and also focus on the up sector instead of the down).   Strictly speaking, all of these models satisfy the MFV hypothesis.    However, as pointed out in Ref. \refcite{Buras:2010mh} if one also requires that the FCNC are suppressed by the third row of $V_{CKM}$ (necessary to suppress the dangerous down-type FCNC), then only the model above will satisfy that requirement.

The couplings to the light Higgs boson can be written explicitly for the symmetry of Eq. (12) (see Botella, et al. \cite{Botella:2015hoa} for a detailed derivation), defining $t_\beta\equiv v_2/v_1$ and $c_{\beta\alpha} \equiv \cos(\beta-\alpha)$ for $i\neq j$ (no sum on k)
\begin{equation}
(N_d)_{ij} = -V^*_{ki}V_{kj} \frac{m_{d_j}}{v} c_{\beta\alpha} (t_\beta + t_\beta^{-1});
\end{equation}   With an obvious extension of the symmetry of Eq. (12), $k$ can be either $d,s$ or $b$, leading to three models.    In the case of Eq. (12):
\begin{equation}
(N_u) = \begin{pmatrix} t_\beta m_u &0&0\cr 0&t_\beta m_c&0\cr 0&0&t_\beta^{-1}m_t\end{pmatrix}
\end{equation}
which clearly shows  FCNC is in the down quark sector.    An alternative to Eq. (12) is
\begin{equation}
Q_{3L} \to e^{i\psi} Q_{3L}, \qquad   d_{3R} \to e^{2i\psi} d_{3R}, \qquad \Phi_2 \to e^{-i\psi} \Phi_2
\end{equation}
which gives
\begin{equation}
(N_u)_{ij} = -V_{ik}V^*_{jk} \frac{m_{u_j}}{v} c_{\beta\alpha} (t_\beta + t_\beta^{-1})
\end{equation}
and
\begin{equation}
  (N_d) = \begin{pmatrix} t_\beta m_d &0&0\cr 0&t_\beta m_s&0\cr 0&0&t_\beta^{-1} m_b\end{pmatrix}
\end{equation}
which gives FCNC in the up quark sector.   Again, $k$ can correspond to $u,c,t$, so there are three models.
A similar symmetry can be used for leptons.       

In many of these models one can relate the couplings to the Cheng-Sher notation\footnote{Strictly speaking, the CS ansatz is not MFV, due to differences in the FCNC couplings of the $Q=2/3$ quarks.}, for example:
\begin{equation}
\lambda_{bs}\frac{(2m_bm_s)^{1/2}}{v} \leftrightarrow \frac{m_b}{v}(t_\beta+t_\beta^{-1})V^*_{ts}V_{tb}X
\end{equation}
where $ X = (\cos(\beta-\alpha), \sin(\beta-\alpha), 1)$ for the coupling to the $h,H$ and $A$, respectively.     Numerically, this gives $\lambda_{bs} = 0.14 (t_\beta + t_\beta^{-1}) X$.   As noted in the text below Table 3, the bound on $\lambda_{bs}$ from $\Delta(M_{B_s})$ is 0.436 for a pseudoscalar mass of 500 GeV and it scales linearly.  It is a factor of three higher for scalars.    One can see that unless $t_\beta$ is fairly large (or very small), this will be completely consistent with experiment.    Thus, BGL models are more consistent with experiment than the pure CS ansatz and they are quite predictive.   The discovery of additional Higgs bosons would yield unambiguous predictions for all flavor-changing couplings.  The decays of the heavier Higgs will be discussed below.

The leptonic sector of BGL models is a little more complicated since the neutrinos can be Dirac or Majorana.    This was first discussed by Botella et al. \cite{Botella:2011ne}.    A more extensive analysis of  the BGL lepton sector can be found in Ref. \refcite{Botella:2014ska}, where many other flavor-changing processes are discussed.    They point out that the mixing angles in the PMNS matrix are much larger and thus Higgs mediated FCNC are not as strongly suppressed.  They also note that in models in which the leptonic FCNC are mainly present in the neutrino sector, the small neutrino masses make it easy to accommodate experimental data.    The BGL approach can be extended to other Higgs structures, such as a 3HDM \cite{Botella:2009pq,Das:2021oik}

\subsection{Flavorful Models}

The bounds on FCNC for first and second generation fields are much more severe.      The CS ansatz and MFV models explain this by small masses and small mixing angles, respectively.    An alternative is to suppose that the source of mass generation for the first two generations is different from that of the third generation.     This was first suggested by Altmannschofer, et al. \cite{Altmannshofer:2015esa} in the context of the apparent CMS measurement (which has now disappeared) of a nonzero $h\to\mu\tau$ decay\footnote{This apparent measurement led to several other papers focused on the MFV hypothesis\cite{Lee:2014rba,He:2015rqa,Baek:2016pef}}.   They considered a 2HDM and another model with a partially composite Higgs.    At about the same time, Ghosh et al. \cite{Ghosh:2015gpa} considered a similar set of models, but motivated by the small first and second generation masses.        Very shortly thereafter, Botella et al. \cite{Botella:2016krk} considered this alternative in the context of the quark sector.   They noted that the 2HDM version gave one of a class of BGL models and studied the phenomenology of this class of models in detail, and then also considered an effective operator analysis involving a TeV completion with vector-like quarks.

\begin{figure}[h!]
\centerline{\includegraphics[width=3.0in]{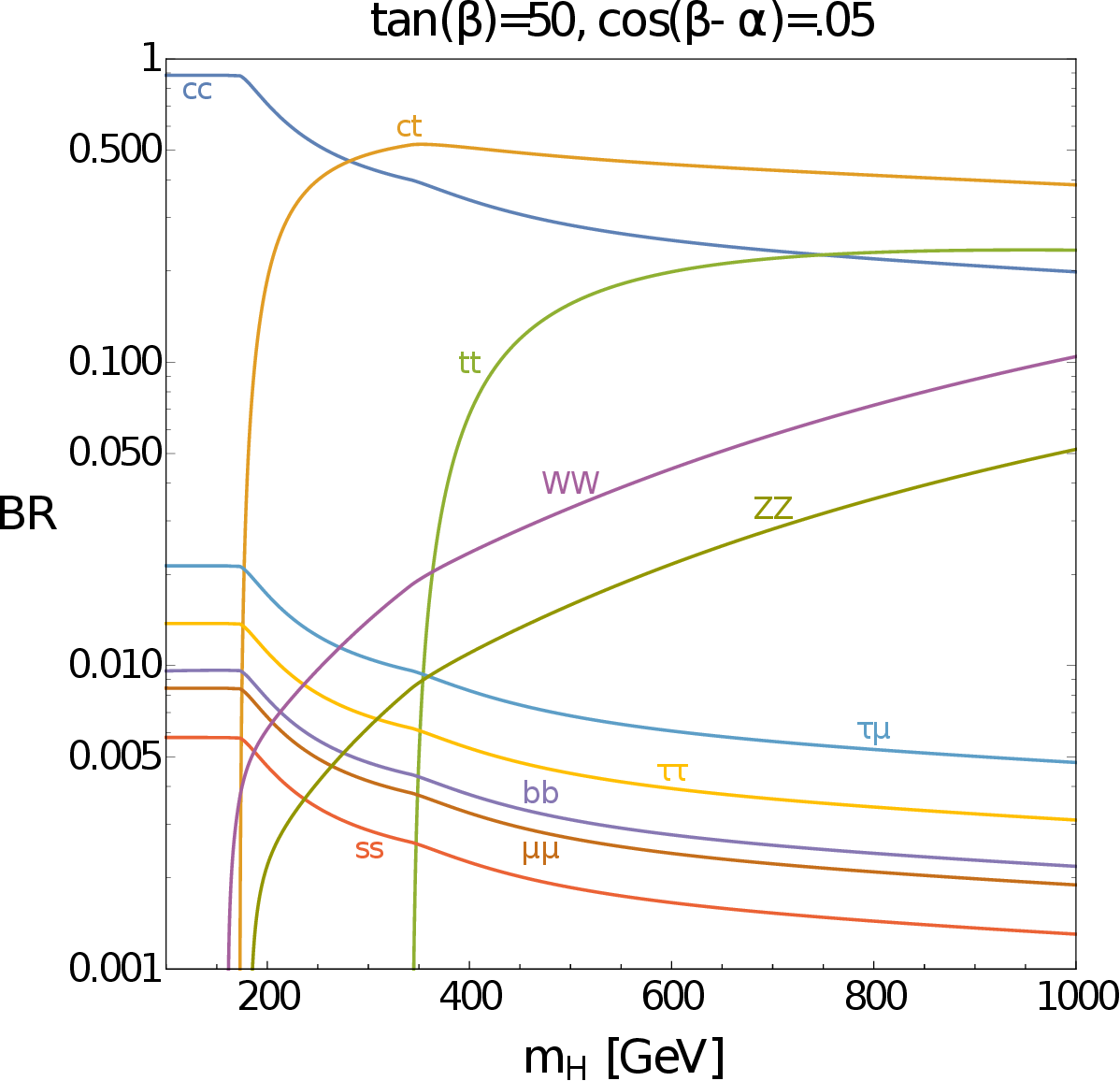}}
\vspace*{6pt}
\caption{The branching ratios of the heavy neutral Higgs as a function of the Higgs mass.   The figure is from Altmannschofer et al. \cite{Altmannshofer:2016zrn}.     Note that the both the hadronic and leptonic decays of the Higgs are dominated by the FCNC decays.     \protect\label{fig3}}
\end{figure}

There was an extensive analysis of the collider phenomenology of the 2HDM version by Altmannschofer et al. \cite{Altmannshofer:2016zrn}.   They noted that the dimuon decay of the Higgs would provide the strongest constraints on deviation from the decoupling ($c_{\beta\alpha}=0$) limit and predicted $h\to\mu\tau$ at the $0.1\%$ branching ratio.   They looked at the leptonic decays of the heavy neutral Higgs and also found that the charmed decays of these Higgs could be substantial, even comparable to top decays.    Figure 3 is from their paper and shows the remarkable fact that reasonable sets of parameters give a heavy Higgs decaying predominantly into $\bar{t}c + \bar{c}t$.  Although this figure explicitly used $\tan\beta =50$, the dominance of the FCNC decay persists down to $\tan\beta=10$.    Charged Higgs decays into the usual $\tau\nu$ and $tb$ are suppressed compared to the usual 2HDMs, showing the importance of looking for other decays.

An alternative approach is to generate the fermion Yukawa coupling by the vev of a general flavon potential.    There are sets of flavors dynamically ``locked" by horizontal symmetries so that each SM quark mass is controlled by a unique flavor.    This could realize the flavorful 2HDM structure in a natural way.   See Ref. \refcite{Altmannshofer:2017uvs} for details.  It turns out that the CKM matrix is automatically hierarchical in this model with $V_{cb}$ and $V_{ub}$ generically of the size observed.    Maddock's PhD thesis \cite{Maddock:2020xtv} ``Theory and Phenomenology of Flavorful Two Higgs Doublet Models" has a very comprehensive discussion of all of these models and an extensive list of references.  It also discusses flavorful models combined with a twin Higgs mechanism \cite{Altmannshofer:2020mfp} as well as rare top decays \cite{Altmannshofer:2019ogm} in these models.

\subsection{Heavy Higgs decays}

As noted earlier, for the couplings of the heavy scalar Higgs boson, the expression is identical to those of the light Higgs with $\cos(\beta-\alpha)$ replaced by $\sin(\beta-\alpha)$ for the coupling of the heavy scalar.  For the couplings of the pseudoscalar, the expression is identical (up to an overall sign) with $\cos(\beta-\alpha)$ replaced by $1$.   
 
Note that since in all models $\cos(\beta-\alpha)$ is small, the flavor-changing couplings of the heavy Higgs bosons can be substantially larger than that of the light Higgs (eqs (16) and (19)) leading to surprising signatures of a heavy Higgs \cite{Sher:2016rhh}.    An updated analysis by Bednyakov and Rutberg \cite{Bednyakov:2018hfq} shows, for example, that the heavy Higgs (scalar or pseudoscalar) can have a branching ratio to $\mu\tau$ of up to 30\% in a BGL type model.    Figure 4 is from their paper and shows that the leptonic decay of the heavy Higgs scalar will always be dominated by the $\mu\tau$ decay, for all values of $\tan\beta$ (the result is similar for the pseudoscalar) .      Gori et al. \cite{Gori:2017tvg} showed in the context of a flavorful model that the {\it dominant} decay mode of a heavy scalar can be into a top and a charm and they discuss production and detection of this mode at the LHC.   They also point out that the dominant decay mode of the charged Higgs, even about the top threshold, can be into charm plus bottom; a study of this process at linear colliders was recently carried out by   Hou et al.\cite{Hou:2021qff}.       In addition, Hou et al. \cite{Hou:2019grj,Hou:2022nyh} studied flavor changing leptonic decays of heavy Higgs bosons at the LHC and future proton-proton colliders.   In fact, CMS has been searching \cite{CMS:2019pex} for lepton flavor violating decays of heavy Higgs bosons in Run II.

\begin{figure}[h!]
\centerline{\includegraphics[width=3.0in]{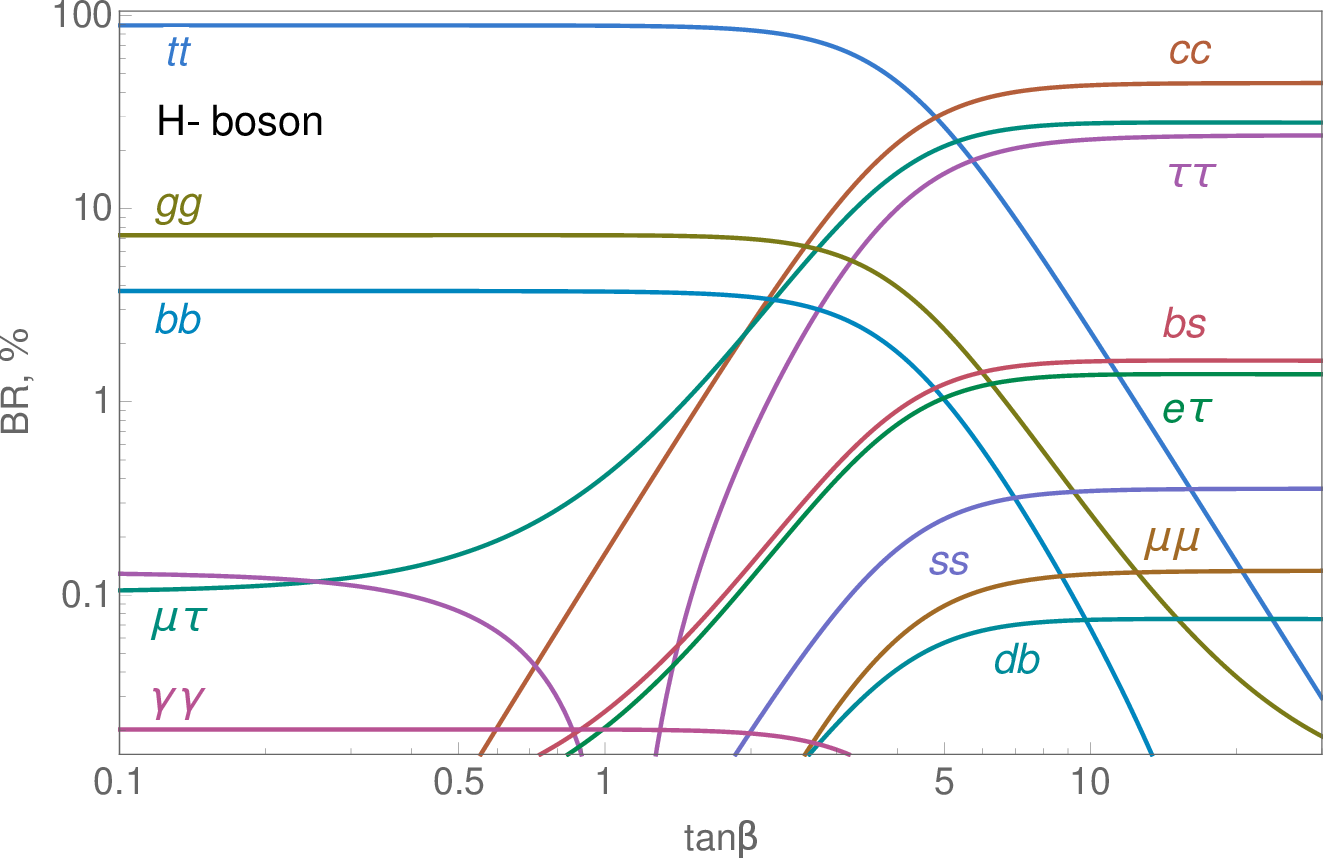}}
\vspace*{6pt}
\caption{The branching ratios of the heavy neutral Higgs as a function of $\tan\beta$ for a Higgs mass of 350 GeV.   It is assumed that $\cos(\beta-\alpha)=0$.   The figure is from Bednyakov and Rutberg  \cite{Bednyakov:2018hfq} .     Note that the both the FCNC leptonic decays of the Higgs dominate.     \protect\label{fig4}}
\end{figure}

\section{Conclusions}

This review has focused on models in which there are tree-level flavor changing neutral currents in the Higgs sector.     After a brief discussion of models without FCNC at tree level, attention was turned to several models with such currents.    The first was a long-standing model based on a mass-matrix ansatz, the so-called Cheng-Sher ansatz in which the flavor changing Yukawa couplings are given by the geometric mean of the Yukawa couplings of the individual fermions.   The ansatz has been used extensively as a guide for experimentalists, but is now being severely challenged by experiments.   Of the nine possible flavor-changing couplings expected by the ansatz to be O(1), five are substantially smaller than 1 (and the other four have not yet been measured).    As a result, the ansatz may no longer be viable.    The focus then turned to the Minimal Flavor Violation hypothesis which posits that the only source of flavor and CP violation arises from the CKM or PMNS matrices.    MFV is a hypothesis not a model and so attention turned to a specific set of models which incorporates the MFV hypothesis, the BGL models.    It was shown that these are more experimentally viable than the Cheng-Sher ansatz but are also extremely predictive - discovery of heavy Higgs bosons will quickly give precise predictions for all of the flavor-changing couplings.    Two other models discussed include flavorful models in which the third generation is treated differently than the other two, and finally flavor changing decays of heavy Higgs bosons were discussed.    The important points of the latter two models can be seen in Figures 3 and 4, in which reasonable models have the dominant decays of heavy Higgs bosons being flavor-changing decays.    This should alert experimentalists to consider such decays in searching for heavy Higgs bosons.

\section*{Acknowledgments}

I am thankful to Tania Robens, Gui Rebelo, Pedro Miguel Martins Ferreira and Gustavo Branco for helpful discussions.   This work was supported by the National Science Foundation grant PHY-2112460.

\end{document}